\documentclass[Afour,sageh,times]{sagej}

\usepackage{float}
\usepackage{moreverb,url}
\usepackage{booktabs}
\usepackage{graphicx}
\usepackage{xurl}
\usepackage[colorlinks,bookmarksopen,bookmarksnumbered,citecolor=red,urlcolor=red]{hyperref}

\newcommand\BibTeX{{\rmfamily B\kern-.05em \textsc{i\kern-.025em b}\kern-.08em
T\kern-.1667em\lower.7ex\hbox{E}\kern-.125emX}}

\def\volumeyear{2026}

\begin{document}

\runninghead{Joy et al.}

\title{Beyond the Hype: Functional, Relational, and Metaphysical Uncertainty in the Pre-Domestication of AI Personal Assistants}

\author{Karen Joy\affilnum{1}, Alyssa Sheehan\affilnum{2} and Tawfiq Ammari\affilnum{1}}

\affiliation{\affilnum{1}Department of Library and Information Science, Rutgers University, New Brunswick, NJ, USA\\
\affilnum{2}Ipsos, Atlanta, GA, USA}


\corrauth{Tawfiq Ammari,
Department of Library and Information Science, Rutgers University,
4 Huntington Street,
New Brunswick, NJ 08901, USA.}

\email{tawfiq.ammari@rutgers.edu}

\begin{abstract}
A new generation of AI personal assistants reached consumers in 2023–2024 amid sweeping claims about anticipatory, agentic intelligence. Wearables such as the Rabbit R1 and Humane AI Pin, and subscription services such as Ohai and Docus, promised to learn users' routines and complete tasks across digital platforms. Drawing on semi-structured interviews with nine early adopters, this article asks how users make sense of these systems when the imaginary of an autonomous "second self" meets the recalcitrance of actual devices. Extending uncertainty reduction theory, we specify three forms of uncertainty in initial encounters: functional (what can it do?), relational (how do I get it to do it?), and metaphysical (what is it to me, and what should it remember?). We find that hype continues the pre-domestication of voice assistants; that the most satisfying uses are user-curated constellations of narrow tools rather than standalone "second selves."
\end{abstract}


\maketitle

\section{Introduction}
A recent wave of AI personal assistants has entered the consumer market: wearable devices such as the Humane AI Pin and the Rabbit R1, and subscription services such as Ohai (calendar management) and Docus (medical guidance). Each was marketed not as an incremental improvement on the smart speakers of the previous decade but as a categorical leap: an agentic ``second self'' that would learn its user's routines, traverse digital platforms, and complete tasks across them. The Rabbit R1's promotional materials promised that a Large Action Model would observe how users interact with the apps they already use and replicate those interactions autonomously \citep{rabbit_research,shepherd_lam_2024}. The Humane AI Pin proposed a screenless, projected interface that would dissolve the smartphone altogether. Both quickly became objects of widespread public disappointment: a viral YouTube review of the Rabbit R1 with more than six million views asked whether the device was a scam \citep{coffeezilla_rabbit_2024}, and Humane abandoned its independent product line not long after launch \citep{wiggers_humanes_2025}.

This article is not primarily an account of two product failures. It is an account of how the early adopters who bought these devices made sense of them: how the cultural imaginary of agentic AI, carried by years of promotional and journalistic discourse, met the brittle, scripted, often unworkable systems that arrived on doorsteps. That meeting is theoretically generative because it isolates a moment that is usually compressed in the history of a successful new medium: the moment at which the prepared meanings of a technology and the technically feasible uses of it visibly diverge, and users must reconcile the two themselves.

\citet{humphry_preparing_2021}, examining the prior generation of smart voice assistants, argue that consumer voice AI was not adopted on the basis of demonstrated utility but \emph{pre-domesticated}: softened in advance through decades of cinematic and literary imagery of conversational machines, and then re-encoded through carefully designed personae (Alexa, Siri, the Google Assistant voice) that translated a long-standing cultural imaginary of dangerous artificial speakers into reassuring domestic ones. The 2023--2024 wave of agentic assistants can be read as another moment of pre-domestication, with the cultural machinery that prepared households for voice-activated information retrieval now being mobilised for autonomous task completion. The cultural framing this time differs in degree rather than in kind: smart speakers arrived with relatively modest advertised utility and accreted relational and creative uses over time as users discovered them \citep{xu_tool_2024,ammari_et_al_19}, whereas agentic assistants arrived with more maximalist advertised utility, and the relational potential their marketing invited has had less room to form before functional difficulties become salient to users.

We pursue this argument through interviews with nine early adopters and through the human--machine communication (HMC) research agenda articulated by \citet{guzman_artificial_2020}, which calls for attending simultaneously to the functional, relational, and metaphysical dimensions of communicative AI. We draw on \citet{xu_tool_2024}'s longitudinal demonstration that functional and relational use of AI voice assistants are not interchangeable and ask what happens when even the functional pathway is itself uncertain. Drawing on uncertainty reduction theory \citep{kramer1999motivation,chang2024uncertainty} and on \citet{pan_humanai_2025}'s analysis of initial human--AI encounters, we extend Guzman and Lewis's framework by specifying, for each of their three dimensions, a corresponding form of uncertainty that early adopters navigate at the initial-encounter moment: \emph{functional}, \emph{relational}, and \emph{metaphysical}. Each maps onto a distinct type of gratification \citep{katz1973uses,wei2024gratification} that participants had been promised, and each was disrupted in ways the existing literature on more mature voice assistants does not anticipate.

\section{Theoretical Framework}

\subsection{From medium to communicator: the HMC tradition}

The communicative AI we examine sits within a longer arc that communication scholarship has been tracing for two decades. \citet{zhao_humanoid_2006} argued, in this journal, that humanoid social robots be understood not as conventional media but as a new \emph{medium} of communication: autonomous, interactive, humanlike entities whose incorporation into social life produces a ``synthetic society in which humans co-mingle with humanoids.'' That argument anticipated a turn that \citet{guzman_artificial_2020} formalised in their human--machine communication (HMC) research agenda: communicative AI cannot be analysed through paradigms that assume human--human communication mediated by passive channels, because the AI itself is the communicator. They propose three analytic dimensions for HMC research: (1) a \emph{functional} dimension, concerning how people make sense of these devices as communicators with particular capabilities; (2) a \emph{relational} dimension, concerning the social bonds users form with and through them; and (3) a \emph{metaphysical} dimension, concerning the ontological work users perform to decide what category of entity they are dealing with.

Each of our three uncertainties takes its name from one of these dimensions. \emph{Functional uncertainty} is the work of figuring out what the AI does as a communicator, and corresponds to the functional dimension. \emph{Relational uncertainty} is the work of figuring out how to address the AI and what register of exchange it will sustain, and corresponds to the relational dimension. \emph{Metaphysical uncertainty} is the work of figuring out what category of entity the AI is (tool, companion, or hybrid), and corresponds to the metaphysical dimension.

\subsection{Pre-domestication and the cultural preparation of voice AI}

The HMC framework specifies what users are doing when they make sense of communicative AI; the domestication tradition specifies the cultural and material setting in which they do it. \citet{silverstone1992information} and \citet{bakardjieva2005internet} argue that household technologies are integrated through a ``moral economy'' of micro-regulations (decisions about who gets access, when, and on what terms), and that these decisions are made against the backdrop of pre-existing meanings that the technology carries into the household.

\citet{humphry_preparing_2021} extend this argument by showing that smart voice assistants entered consumer households already laden with cultural meanings drawn from a long history of cinematic AI (HAL, Samantha, the ship's computer), and that manufacturers leveraged these inherited imaginaries through carefully chosen voice personae, naming conventions, and invocation rituals. They call this \emph{pre-domestication}: the cultural work that prepares a technology for the home before any specific household integrates it. The first cycle of pre-domestication, around 2014--2018, produced smart speakers whose advertised capabilities were modest (``ask me what the weather is'') and whose relational and creative uses emerged gradually as households discovered them \citep{porcheron_voice_2018,ammari_et_al_19,purington_alexa_2017}.

The current wave shifts this dynamic. Agentic AI assistants have arrived with maximalist advertised capabilities (autonomy, anticipation, agency) and with cultural framings that draw less on benign cinematic AI than on the marketing of disruptive Silicon Valley products. This shift in the rhetorical framing of pre-domestication is consequential: it raises the expectation gulf \citep{luger_like_2016} that users must traverse and sets up the functional collapse we document below. Recent empirical work extends the domestication frame to generative AI  \citep{ammari2026learning}, showing that functional competence with the assistant emerges not from a single moment of adoption but from ongoing relational negotiation across distinct use genres (e.g., academic workhorse, emotional companion, metacognitive partner). This work examines an earlier moment in that trajectory, the initial encounter, and focuses on a class of device for which the pre-domestication \citep{saariketo2018unchallenged} has run further ahead of technical feasibility than was the case with chat-based generative AI.

\subsection{Functional and relational use, and the gratifications they support}

\citet{xu_tool_2024}, the most direct empirical predecessor to this study, distinguish the \emph{functional use} of AI voice assistants (issuing commands, retrieving information) from their \emph{relational use} (small talk, self-disclosure, treating the assistant as a social being). Their two-wave panel of 354 users yields a counter-intuitive finding: functional use \emph{reduces} subsequent relational use through a corresponding reduction in self-disclosure, while relational use reinforces itself over time. Voice assistants, in short, do not glide naturally from tool to companion; the two pathways are partially exclusive. We extend this argument to an earlier point than Xu and Li's panel captured: the initial encounter, when functional use is itself unstable. \citet{pan_humanai_2025} show experimentally that AI agency shapes early trust and liking judgments, which then organise subsequent willingness to engage. Our participants are at this initial-encounter moment, and we observe that when the functional pathway collapses, the relational pathway is foreclosed before it can begin.

Uses-and-gratifications scholarship \citep{katz1973uses,rubin2009uses,wei2024gratification} identifies four types of gratification that drive adoption of communicative technologies, which we use to specify what each of our uncertainties disrupts. \emph{Technological} gratification is satisfaction from engaging with a novel or advanced technology, often tied to identity and status \citep{mclean2019hey}. \emph{Utilitarian} gratification is the value of effectively completing tasks (scheduling, retrieval, device control). \emph{Hedonic} gratification is pleasure from casual, playful, or creative use \citep{jo2022continuance}. \emph{Social} gratification covers companionship, perceived relational connection, and the sense that the device occupies a recognisable social role \citep{SocialContextUGT}. A fifth UGT category, \textit{conversational repair}, captures how users persist through breakdowns via prompt adaptation and clarification rather than abandoning the interaction \citep{ammari2025students}; in our case it becomes a central response to relational uncertainty.

\subsection{Uncertainty reduction in initial human--AI encounters}

\citet{kramer1999motivation} and \citet{chang2024uncertainty} frame technology adoption as a process of managing ambiguity through both \emph{interactive} reduction strategies (engaging the system directly, re-prompting, testing limits) and \emph{passive} ones (watching others, reading reviews, attending to media coverage). \citet{zhang2022uncertainty} note that with LLM-powered systems the surface of unfamiliarity is broader than with previous interfaces: users are uncertain not only about features but about the underlying epistemics of the system, that is, what it knows, what it remembers, what it is for. Building on this, and extending \citeauthor{guzman_artificial_2020}'s (\citeyear{guzman_artificial_2020}) HMC research agenda, we specify a corresponding form of uncertainty for each of their three dimensions, recurring across our participants' accounts of initial encounters:

\begin{enumerate}
    \item \textbf{Functional uncertainty} (\emph{functional} HMC dimension) is uncertainty about what the assistant can reliably do as a communicator. It arises when system boundaries are ambiguous, when promoted features fail in practice, and when users cannot predict which tasks will succeed. It primarily disrupts \emph{technological} and \emph{utilitarian} gratifications.
    \item \textbf{Relational uncertainty} (\emph{relational} HMC dimension) is uncertainty about how to address the assistant effectively and about the register of exchange it will sustain. It arises when the device's input style is unclear (what tone? what command form? what level of specificity?) and when system responses do not reliably signal whether the user has been understood. It primarily disrupts \emph{hedonic} gratifications and, where users sought deeper relational engagement, \emph{social} gratifications.
    \item \textbf{Metaphysical uncertainty} (\emph{metaphysical} HMC dimension) is uncertainty about what category of entity the assistant is (tool, companion, or hybrid) and about what it is reasonable to share with it, expect of it, or treat it as remembering. It primarily disrupts \emph{social} gratifications.
\end{enumerate}

These three uncertainties are not exhaustive, and they overlap at their edges (relational uncertainty bleeds into functional uncertainty when users cannot tell whether a failed interaction was a bad prompt or a missing capability). But together they specify where the sociotechnical gap \citep{AckermanGap}, the disjunction between socially situated needs and technically feasible implementations, becomes locally visible to users in initial encounters with agentic-AI assistants.

\section{Method}

We conducted semi-structured interviews (60--90 minutes) with nine early adopters of Rabbit R1, Humane AI Pin, Ohai, or Docus between mid-2024 and late 2024. Recruitment was through targeted outreach to verified owners and subscribers via online communities (\href{https://www.reddit.com/r/RabbitR1/}{r/RabbitR1}, \href{https://www.reddit.com/r/AiPin/}{r/AiPin}) and personal networks; we required active ownership of the device or active subscription to the service in order to ground the interview in current experience. The protocol covered daily routines, prior technology use, the path from advertising exposure to purchase, initial set-up, integration with existing tools, and concerns about data and privacy. Interviews were conducted by video conference, recorded with consent, and transcribed verbatim. The study was approved by the authors' institutional ethics review.

Analysis proceeded inductively in two passes: open coding of the first wave of transcripts produced an initial set of themes (minimising distraction, personalisation, distribution of labour, mistake-proofing), which we then refined against the human--machine communication and uncertainty literatures into the three-part typology presented below. We followed a hierarchical, team-based approach to qualitative analysis \citep{cascio2019team}. The first author conducted line-by-line open coding of each transcript and then discussed the resulting codes and emerging themes iteratively with the rest of the team, working toward consensus on the coding structure and on contested or ambiguous excerpts. Coding was performed iteratively, with memo-writing between interviews; saturation on the central themes was reached by the eighth interview.

Participants ranged in age from the late twenties to the late fifties; all identified as men. Seven were Rabbit R1 owners, two had purchased the Humane AI Pin, two subscribed to Ohai, and one to Docus (several had multiple devices/subscriptions). All had prior experience with at least one legacy voice assistant and with text-based generative AI (most commonly ChatGPT). The male-only sample is a meaningful limitation: research on AI adoption finds a pronounced gender skew among early adopters, but the skew is not absolute, and we return to this in our limitations. Table~\ref{tab:demographics} in the Appendix summarises the sample.
 
We use P1--P9 to identify participants. Direct quotations are reproduced from transcripts; minor disfluencies have been silently removed where they did not bear on meaning.

\section{Findings}

\subsection{Hype as cultural preparation: pre-domestication in the YouTube era}
\label{sec:findings1_hype}

\citet{humphry_preparing_2021} show that smart voice assistants were prepared for consumer households through decades of cinematic and literary imagery, then re-encoded by manufacturers through deliberate voice personae and invocation rituals. The 2023--2024 wave of agentic assistants underwent a comparable preparation on a shorter timescale, carried less by half a century of science fiction than by a concentrated few months of viral promotional video, influencer coverage, and tech-press anticipation. The infrastructure of YouTube is a feature of this moment worth flagging: pre-launch keynotes, reaction videos, and creator commentary circulated at high volume in the run-up to release, providing much of the cultural envelope into which our participants placed the new devices, and (as we note below) the same infrastructure produced the influential review videos that later questioned them. We do not treat this YouTube-mediated pre-domestication \citep{saariketo2018unchallenged} as a finding in its own right, but it is the medium through which the cultural preparation described by \citeauthor{humphry_preparing_2021} now travels.

Participants described their purchase decisions in terms that recognise this preparation explicitly. P2, who bought both a Rabbit R1 and a Humane AI Pin, recalled: ``man, the advertisement, the YouTube videos, the Ted Talks, all that stuff was just like, whoa, really? That's crazy for \$199; you can't lose, you know what I'm saying?'' P3 said of the Humane Pin: ``Ambient computing for the real world\ldots and it looks sophisticated. When I wear it I honestly feel it shows professionalism.'' P4 expected the Rabbit to act on intent the way a competent assistant would: ``I thought it could have a very clear sense and just do it---like, have the `action' part to it\ldots like order me a pizza\ldots and carry out the task\ldots based on my past order(s).'' What the marketing prepared participants for, in other words, was not a particular feature set but an entire \emph{ontological category}: a device that would act on their behalf in the agentic sense.

The functional collapse that followed was therefore not a discovery of bugs but a discovery that the prepared category was empty. P2, on the Rabbit's much-publicised Large Action Model: ``It indicated that it was a learning LAM, some type of learning module in which it would get the information and do all these tasks for you. It's not able to do that. That technology doesn't even exist.'' P6 documented the brittleness with a kind of forensic patience: ``I will mess with [the Rabbit] logic, and it'll do the same thing\ldots that's a pre-programmed response. I'll even say things like play me anything, excluding the Beatles\ldots and it'll still play the Beatles.'' Importantly, P6 had already absorbed the discount that comes with experienced adoption of hyped tech: ``I knew it was all marketing. I knew it was never going to be what they promised. I'm not fully disappointed.'' P4, with less practice in this discount, was looking forward to a class-action lawsuit and described the Rabbit as ``a clicky [McDonald's] happy-meal toy [or] a Tamagotchi that speaks.''

Two analytic observations follow. First, the rhetorical framing of this wave of pre-domestication (autonomy, agency, action) sits somewhat differently from the helper framing of the prior generation. Smart speakers were largely marketed as helpers; agentic assistants are marketed as proxies. The expectation gulf \citep{luger_like_2016} is correspondingly wider, and the tenor of failure is correspondingly more affectively charged: not ``it didn't catch what I said'' but ``it's a scam.'' Second, where the pre-domestication that stabilised voice AI in the home drew on a broad ecosystem of cultural objects (films, songs, jokes about Alexa), this wave's cultural preparation leans more heavily on promotional video and creator commentary on platforms like YouTube. The framing artefacts here are concrete and traceable. Rabbit's CES 2024 launch keynote, in which CEO Jesse Lyu walked through agentic demos of ride-hailing, food ordering, and trip planning, accumulated more than 4 million YouTube views within its first week of release and has continued to circulate well beyond that \citep{rabbit_keynote_2024}. Imran Chaudhri's TED talk previewing the Humane AI Pin, ``The Disappearing Computer,'' has been viewed roughly 2.3 million times on TED's own platform \citep{chaudhri_ted_2023}. These were not incidental promotional artefacts; they did much of the cultural work of preparing the category. That medium cuts both ways: the same channels that prepared the category produced, within months, the six-million-view review by Coffeezilla \citep{coffeezilla_rabbit_2024} that pulled it apart.

Some participants compared the new devices to legacy assistants and found, paradoxically, that the legacy systems performed the social work better. P5 and P6 noted that the next-gen devices produced more elaborate responses than Siri or Alexa, but P3 and P2 found their error-handling clumsy: raw system messages, dropped calls, broken media controls. \citet{nass_can_1995} argued thirty years ago that people respond socially to computers that perform social roles competently; legacy assistants, through long iteration, have learned the small repairs that sustain that performance, and the next-gen devices have not.

Relational uncertainty was the most variable across our sample. P3 articulated the mismatch most clearly in his account of the Humane AI Pin, the screenless wearable pin that was pitched as a replacement for the smartphone: the device's hardware design carried an aesthetic of sophistication, but the interface (a gesture vocabulary projected onto the palm, paired with a small set of voice invocations) gave him no reliable way to issue commands or retrieve information. The sophistication of the object signalled a competence the interaction could not deliver. For P4, the same uncertainty surfaced on the hedonic side. When he reached for the Rabbit to do casual storytelling or to bounce playful prompts, the device returned what he described as generic or off-topic filler, foreclosing the entertainment role he had wanted from it. P2, having paid for the Humane Pin and tried to integrate it into daily wear, eventually returned it: the projected display failed reliably enough in bright or uneven environments that the device's claim on a place in his day did not survive ordinary outdoor use.

The relational finding was not uniformly negative, however. P7 took the opposite reading of the Rabbit's interface and built a stable niche use around it. For him, the device's most useful feature was its low ceremony: a single push-button voice invocation, no biometric authentication, no screen wake. He used it during his commute, where Siri's repeated demand for Face ID authentication was a problem the Rabbit simply did not pose. ``I don't want to do face ID and all that when I am driving,'' he said. ``With Rabbit, I can just push the button and go.'' The same brittle interface that gave P3 and P4 nothing gave P7 something quite specific: a hands-free, low-friction voice capture instrument scoped tightly to the car. Relational uncertainty, in other words, was not a uniform property of these devices but a property of the fit between the device's affordances and the particular activity context the user tried to deploy it in.

\subsection{Bricolage in place of the ``second self'': value as constellation}
\label{sec:findings2_constellations}

The most consistent finding across our interviews is that the participants who reported the highest satisfaction were the ones who had stopped expecting their agentic assistant to be a stand-alone ``second self'' and had instead pieced it into a personally curated constellation of tools, each used for the one narrow task it actually did well.

P6, an emergency-room nurse with ADHD, built the most articulated version of this bricolage. The Rabbit R1, a pocket-sized handheld operated through a single push-to-talk button, had headline features---the Large Action Model that was meant to operate apps on the user's behalf, the integrations with Uber and Spotify---that were useless to him. But the device's voice-recording function was robust, and ChatGPT's summarisation was robust, and together they let him do something neither system advertised: ``The joke is that I have 30 seconds to [see the patient] and 30 minutes to chart it\ldots Now I can do 12 [patients] within an hour\ldots easily, easily!'' He recorded his patient encounters on the Rabbit, uploaded the recordings to ChatGPT, and produced charting drafts. ``Even just the mental energy alone\ldots is incredible.''

P1, a director of business operations, used Ohai (a calendar service trained on natural-language scheduling) in a similar narrow-and-deep way: ``I just text: `I have a sales meeting tomorrow from 7 to 8:30 a.m.\ldots' because I know Ohai can understand it and execute what I'm trying to get it to do.'' He explicitly contrasted Ohai with the general-purpose assistants: ``Ohai kind of bridges the gap between one of the purely generative AI models\ldots and the integrations you would get with a workspace on Gemini.'' P5 used Docus (a medical-guidance service) for differential-diagnosis conversations and rated it about ``50/50'': imperfect, but useful in the specific role he gave it.

The narrow scoping was not, however, a guarantee of satisfaction. P5's use of Docus had a recursive failure mode that the more general-purpose tools did not exhibit. The service could conduct reasonable differential-diagnosis conversations, but it would refuse drug-interaction queries and route him back to a human clinician instead. ``It's not very useful,'' he said, ``if you ask the AI a question and end up getting recommended to consult your doctor again.'' The recursion is its own particular failure: a task-specific service that was supposed to reduce his clinical search burden was returning him to that burden by referring him back to the very expert consultation he had subscribed in order to bypass. The narrow scoping that protected against hallucinated answers on dangerous medical questions also protected against the service being useful for the questions it had been engineered to handle.

A second bricolage pattern operated outside the medical and scheduling cases. P2 used the Rabbit's voice-prompt pipeline to chain into Midjourney for visual content production, generating imagery for a mental-health-awareness social-media campaign he was running. The Rabbit's image-generation plumbing on its own would have been an unstable foundation for the workflow; what made it work was that the device was passing prompts to a separately-curated image service P2 had already vetted. The constellation logic was the same as P6's charting workflow: the agentic assistant was useful insofar as it was a voice-input layer in front of tools the user trusted independently.

These uses look like what \citet{forlizzi2008product} calls a \emph{product ecology}: the value of any one device is not in its features but in the role it plays in a configuration of other devices, people, and routines that the user assembles. Two further observations sharpen this point. First, the agentic-AI marketing pitched these devices as \emph{ecology terminators} (the Humane Pin was to replace the smartphone, the Rabbit to subsume the apps), but participants used them only when they could be \emph{ecology participants}, subordinate elements in a constellation the user controls. Second, the bricolage requires technical fluency that not all participants had. P6's nurse-charting workflow is real but non-trivial to assemble; users without his comfort moving recordings between platforms cannot duplicate it. The shift from ``second self'' to ``constellation'' thus redistributes labour from the device manufacturer to the user, and unequally.

This finding maps onto \citet{xu_tool_2024}'s functional/relational distinction in a specific way. The constellation users were achieving high \textit{utilitarian} and \textit{technological} gratifications by deliberately scoping the assistant to a narrow tool role. They were not pursuing relational use; indeed, several explicitly rejected it. P1, when asked whether he wanted Ohai to feel more like a companion, said: ``Organizing, organizing, organizing\ldots work schedule, home schedule\ldots no conflicts.'' What our case adds to Xu and Li's panel result is the earlier timeline: the narrowing-to-tool can be deliberate and conscious from the first encounter, not just an emergent outcome of accumulated functional use.

\subsection{Memory, role, and the multidimensional negotiation of privacy}
\label{sec:findings3_metaphysical}

Where the constellation users had narrowed their assistants to tools, other participants wanted the opposite: assistants that remembered them, knew their context, and could carry conversation across sessions. P9's case is the clearest illustration in our data of an attempted relational pathway. He treated the Rabbit, a handheld voice-capture device, as a kind of walking biographer, dictating reflections during his daily walks and trusting that a searchable record would remain. The RabbitHole feature, which logged past interactions in a cloud-side archive, was the affordance that made this use legible to him: ``If I want to go back again, I have the whole script of the conversation.'' But the relational coherence he hoped for, in which the assistant carried earlier walks forward into the texture of later ones, was foreclosed by the device's session-by-session amnesia. The archive was retrievable but inert: it did not shape what the device said to him on the next walk. P5 imagined an alternative version of Docus that would carry his context forward in exactly this way: ``It'll know me so well that it'll say, `Hey, you haven't spoken to your uncle in Cuba in a week. Want me to mark him on your calendar?' I'd pay for that.''

For these participants, the principal frustration was not a functional collapse but a \emph{metaphysical} one: the assistant did not maintain continuity across sessions, did not remember preferences, did not behave like a partner who had been there yesterday. P6 articulated this most clearly: ``I can't tell it to remember things\ldots Not really. Does it remember my preferences? No.'' The labour of maintaining continuity therefore fell on the user, who had to re-establish context at the start of each session, a labour that P6 noted he was now performing with humans as well: ``I convey my needs [to Rabbit] so much that now I do it with people too.''

The metaphysical question \citep{guzman_artificial_2020} of what kind of entity is on offer (the question that, in our framework, surfaces as metaphysical uncertainty) was therefore unsettled in a way that the relational research on legacy voice assistants does not capture. \citet{purington_alexa_2017} found that owners of Amazon Echos personified the device, talked to it casually, and developed durable patterns of social address. Our participants, by contrast, were oscillating between social modes session by session, because the device did not retain enough state to anchor a stable role. \citet{xu_tool_2024} showed that relational use of voice assistants depends on self-disclosure, and self-disclosure depends on the perceived continuity of the listener. When that continuity is technically absent, relational use cannot accrete, regardless of what users initially wanted from the device.

This bears on the privacy question in a way that the existing literature on smart-speaker privacy does not always anticipate. \citet{nissenbaum2004privacy} defines privacy as \emph{contextual integrity}: the appropriate flow of information within a particular social setting. \citet{kang_communication_2023} extend this argument by specifying three distinct strategies users deploy to manage privacy with smart speakers: \emph{disclosure} (what to share at all), \emph{boundary linkage} (who else gets access through the device), and \emph{boundary control} (over what the device retains and how). Crucially, they show that privacy self-efficacy (the user's sense of being able to act effectively on these strategies) predicts engagement across all three.

In this light, the ``privacy apathy'' that several of our participants displayed reads differently than a simple indifference. P4 joked that he did not mind data collection: ``I get the need for privacy\ldots but I don't care too much. Are you gonna give me more targeted ads? Cool, thank you.'' P5 was unconcerned about why Docus did not recall his medical history. These positions are not absence of privacy logic; they are positions taken \emph{within} a multidimensional privacy logic that recognises the trade-off between disclosure and personalisation and resolves it differently than a more privacy-active user would. What participants were missing was less concern than \emph{self-efficacy}: the perceived ability to act on boundary control. ``I can't tell it to remember things\ldots Not really'' (P6) is a statement about boundary control, not about disclosure. The assistants did not give users a meaningful interface for negotiating context fidelity, and so even users who would have wanted to negotiate it could not.

Older adults using voice assistants as companions have been documented sharing sensitive medical and financial information with devices they perceived as friends \citep{oewel_et_al_23}.  \citet{huang2025he} similarly shows that users adopt AI as a private channel of communication when human channels feel exposed. Our participants were younger, more technically fluent, and yet they were navigating the same boundary-control question without adequate tools. The implication is that contextual-integrity questions are not a niche concern raised by particular vulnerable populations; they are the structural form that metaphysical uncertainty takes for any user who is being asked to decide what kind of entity is on the other end of the conversation.

\section{Discussion}

Our typology extends \citet{guzman_artificial_2020}'s three-dimension HMC framework by identifying, for each dimension, a corresponding form of uncertainty that early adopters navigate at the initial-encounter moment: functional uncertainty in the functional dimension, relational uncertainty in the relational dimension, and metaphysical uncertainty in the metaphysical dimension. Each disrupts distinct types of gratification that the marketing of these devices had primed. Functional uncertainty disrupts technological and utilitarian gratifications: users who bought into the imaginary of agentic action find the action absent. Relational uncertainty disrupts hedonic and (where users sought it) social gratifications: even when the device works, users cannot tell what mode of address it expects. Metaphysical uncertainty disrupts both the relational pathway \citet{xu_tool_2024} identifies and the boundary-management work \citet{kang_communication_2023} documents: users do not know what social category the assistant occupies, and they have no fine-grained tools to negotiate that occupation. Table~\ref{tab:uncertainty_gratification} summarises how each form of uncertainty manifested across the four gratification types in our data.

\begin{table*}[h]
\small
\centering
\caption{How the three forms of uncertainty disrupt distinct gratification registers \citep{katz1973uses,wei2024gratification}. Cells contain the empirical manifestation observed in our participants; ``---'' indicates no consistent manifestation in that cell. The diagonal pattern is read-able as evidence that functional, relational, and metaphysical uncertainty are partially distinct phenomena rather than collapsible into a single dimension of system unreliability.}
\label{tab:uncertainty_gratification}
\begin{tabular}{@{}lp{2.9cm}p{2.9cm}p{2.9cm}p{2.9cm}@{}}
\toprule
& \multicolumn{4}{c}{\textbf{Gratification}} \\
\cmidrule(l){2-5}
\textbf{Uncertainty} & Technological & Utilitarian & Hedonic & Social \\
\midrule
Functional      & Advertised AI capabilities did not exist & Could not complete promised tasks; users assembled tool constellations & Playful features felt limited or broken & --- \\
\addlinespace
Relational   & Hardware/interface clashed with sophisticated framing & --- & Unclear how to prompt engaging or fun output & Continuity broke across sessions; relational mode unstable \\
\addlinespace
Metaphysical          & --- & --- & --- & Assistant could not occupy a stable companion role; boundary-control affordances absent \\
\bottomrule
\end{tabular}
\end{table*}

The empty cells in Table~\ref{tab:uncertainty_gratification} are themselves analytically important. Metaphysical uncertainty does not disrupt utilitarian gratification in our data: users who scoped their assistants narrowly to tool roles (Finding~2) were not troubled by the assistant's unsettled ontological category, because they had pre-emptively settled it as ``not-a-companion.'' This is the initial-encounter version of the foreclosure that \citet{xu_tool_2024} document longitudinally: functional and relational use trade off against each other, and the trade-off can be made at first contact.


The earlier failures of smart speakers (mis-hearings, dropped commands, awkward Alexa jokes) were largely \emph{functional} failures within a well-prepared cultural envelope \citep{humphry_preparing_2021}. Relational uses accreted alongside them because the cultural envelope still held \citep{xu_tool_2024}. The failures we observed look more like strains on the cultural envelope itself: the agentic-AI imaginary that prepared the Rabbit and the Humane Pin for consumer households is not yet stabilised by the products themselves, and the relational and metaphysical pathways have had less room to settle than they did with smart speakers. The users who got the most value used the device as one element within a tool ecology of their own making, scoping it to the narrow tasks it could reliably perform alongside other tools they already trusted.

Our case has implications for the sociotechnical-gap argument \citep{AckermanGap}. The gap is usually described as a structural disjunction between socially situated needs and technically feasible implementations, to be narrowed by better design. We identify a second source: the gap can be widened by cultural preparation that runs ahead of technical feasibility. When pre-domestication promises a category of entity that the technology cannot yet deliver (an autonomous agentic assistant or a second self), even technically competent implementations fail to be socially competent, because the expectations they must meet are calibrated to a device that does not yet exist. Closing the gap is therefore not only a design problem but a rhetorical one: the framing under which a technology arrives is part of what its design has to satisfy.

A second implication is theoretical. Building on Xu and Li's finding that functional use can suppress self-disclosure and forestall relational deepening in mature voice-assistant use, we observe a parallel dynamic at the initial-encounter moment \citep{pan_humanai_2025}: when functional use is unstable and users adopt a deliberately narrow scoping strategy, the relational pathway is closed off from the start. The HMC dimensions \citep{guzman_artificial_2020,zhao_humanoid_2006} should therefore be read not as parallel tracks along which users move independently but as competing settlements that an initial encounter may push toward one or the other.

A third implication concerns privacy. The dominant narrative of ``privacy apathy'' among AI users treats indifference to data collection as a moral or cognitive failing. \citet{kang_communication_2023}'s multidimensional framework allows a more generous and more accurate reading: apparent apathy is often a particular settlement of the disclosure/boundary-linkage/boundary-control trade-off under conditions of low boundary-control self-efficacy. If the device does not let you negotiate what it remembers, you cannot act on a preference about what it remembers, and the position that looks like apathy is what acting on the only available lever (overall disclosure) actually looks like.

\section{Implications}

The cultural preparation that domesticated smart voice assistants is being mobilised again for agentic AI assistants that the underlying technology cannot yet fully produce. The early adopters who bought the first wave of these devices encountered three forms of uncertainty (about what the device does, how to address it, and what kind of entity it is) that the previous generation of voice-assistant research, focused on stable household integration, does not fully anticipate. The users who got the most out of these devices stopped expecting them to be the second selves they had been marketed as and started using them as constellation elements in tool ecologies of their own design. The users who wanted the relational pathway found it foreclosed before it had a chance to begin, because the devices could not maintain the continuity of context that relational use requires. And the privacy ``apathy'' observable in the sample is better read, in the multidimensional terms \citet{kang_communication_2023} provide, as a particular settlement of the trade-offs available when boundary-control self-efficacy is low. The sociotechnical gap \citep{AckermanGap} between marketed and lived AI assistants is not narrowing under the pressure of new product launches, and the cultural machinery of pre-domestication is part of why it is not.

\section{Limitations and future work}

Three limitations bound our argument. First, the sample is small and all-male. Research on AI adoption documents a pronounced gender skew among early adopters, and our sample reflects this; we make no claim that the dynamics we observe generalise across gender, and a parallel study with women adopters would clarify which findings are about agentic-AI initial encounters in general and which are about a particular slice of the early-adopter population. Second, the sampling frame excludes users who returned their device or cancelled their subscription, producing a survivor-bias problem familiar to qualitative work on emerging technology: the users we did not interview may have experienced the functional collapse more sharply than the ones we did. Third, the interview moment captures an initial-encounter window. Whether the foreclosure of the relational pathway we document persists or whether some participants accrete relational use over time as the devices mature is an open longitudinal question, one for which \citet{xu_tool_2024}'s two-wave panel design offers an obvious model.

\bibliographystyle{SageH}
\bibliography{GapNMS}

%

%
\clearpage
\appendix
\onecolumn
\section{Participant details}

\begin{table}[H]
\small
\centering
\caption{Participants, their AI-assistant adoption, and prior technology use. Race is reported as participants self-described; compound entries reflect self-described multi-racial identity.}
\label{tab:demographics}
\begin{tabular}{@{}ll p{4.0cm} l p{1.7cm} p{2.8cm} p{2.6cm}@{}}
\toprule
\textbf{Participant} & \textbf{Age} & \textbf{Prior technology used} & \textbf{Gender} & \textbf{Race} & \textbf{AI assistant(s) adopted} & \textbf{Job/title} \\
\midrule
P1 & 36--45 & Smart speakers; virtual assistants; smart-home devices; fitness trackers; text generators; image generators & Man & White & Ohai (subscription service) & Director of Business Operations \\
\addlinespace
P2 & 46--55 & Smart speakers; virtual assistants; smart-home devices; fitness trackers; text generators & Man & Black/Latino & Rabbit R1 and Humane AI Pin (devices) & Clinical Supervisor \\
\addlinespace
P3 & 26--35 & Smart speakers; virtual assistants; smart-home devices; text generators; image generators & Man & White & Humane AI Pin (device); Ohai (subscription service) & Portfolio Analyst \\
\addlinespace
P4 & 26--35 & Virtual assistants; text generators; image generators; video generators & Man & Black & Rabbit R1 (device) & Manager of Partnerships \\
\addlinespace
P5 & 36--45 & Smart speakers; virtual assistants; smart-home devices; text generators; image generators; video generators; audio generators & Man & Latino & Docus (subscription service) & AIM Program Manager \\
\addlinespace
P6 & 26--35 & Smart speakers; virtual assistants; smart-home devices; text generators; image generators; video generators; audio generators & Man & Latino/Native & Rabbit R1 (device) & Nurse \\
\addlinespace
P7 & 55+ & Smart speakers; virtual assistants; smart-home devices; text generators; image generators; audio generators & Man & Black & Rabbit R1 (device) & Retired \\
\addlinespace
P8 & 26--35 & Smart speakers; virtual assistants; smart-home devices; text generators; image generators & Man & Asian & Rabbit R1 (device) & Analyst \\
\addlinespace
P9 & 26--35 & Smart speakers; virtual assistants; smart-home devices; text generators; image generators & Man & Asian & Rabbit R1 (device) & Software Analyst \\
\bottomrule
\end{tabular}
\end{table}


\end{document}